\documentstyle[epsfig,graphicx,12pt]{article}%
\begin{document}%
\centerline{\large\bf Production of Pseudoscalar Mesons}%
\centerline{W.J. Briscoe, R.A. Arndt, I.I. Strakovsky and
R.L.Workman}%
\centerline{\small \it The Center for Nuclear Studies and
Department of Physics} %
\centerline{\small \it The George Washington University}%
\centerline{\small \it Washington, D.C. 20052}%
\medskip%
\centerline{\small \bf Abstract}%
Experiments that study the hadronic and electromagnetic production
of the pseudoscalar mesons -- pions, etas and kaons, contribute to
our knowledge of the properties of baryon and hyperon resonances.
Fixed-target programs at hadronic facilities such as BNL-AGS have
been phased out.  However, the availability of modern experimental
facilities with pseudo-monochromatic or tagged medium-energy
photon beams at GRAAL, SPring-8, Bonn, Mainz, and Jefferson Lab,
together with LEGS, Max-Lab, and HIGS at lower energies, are
beginning to produce high-quality results. These new data have
smaller statistical uncertainties and better understood systematic
uncertainties, than those obtained at the older bremsstrahlung
facilities, for measurements of differential and integrated cross
sections, as well as polarization and asymmetry. Experimental
results are compared with the predictions of QCD-based approaches,
such as the lattice-gauge calculations of baryon properties, and
Chiral Perturbation Theory applied to threshold photoproduction,
and are essential to the performance of Partial-Wave Analyses
(PWA). These PWA studies are less model dependent than in the
past, and are used in coupled-channels calculations that
incorporate unitarity dynamically, and combine results from
hadronic reaction channels with electromagnetic processes. This
approach is necessary to extract resonance properties and may lead
to the discovery of the ``missing resonances" predicted by a
number of different QCD-inspired calculations. We discuss recent
experimental and phenomenological results for single and double
pseudoscalar meson hadronic and photoproduction channels with
emphasis on the JLab Hall B and the BNL/AGS Crystal Ball programs.

\newpage

\section{Introduction}

In the following, we review a set of experiments, utilizing both
electromagnetic and hadronic probes, designed to enhance our
understanding of the baryon and hyperon resonances. These programs
are complementary, since hadronic information (mass, width, and
branching fraction) is required in the extraction of photo-decay
amplitudes. The overall program is quite broad, aiming not only to
improve upon existing estimates (obtained mainly from single-pion
photoproduction and $\pi N$ elastic scattering), but also
including the production of other mesons (such and the $\eta$ and
$K$), and including multi-particle final states.\cite{ra02} These
processes are more difficult to analyze, but must be included if
we are to solve the missing resonance problem and provide
stringent constraints for QCD-inspired models and lattice
calculations.\cite{sd03}

\section{Photoproduction of Pseudoscalar Mesons at Jlab}

The first round of real-photon experiments using the Tagged Photon
Facility~\cite{job00} and CLAS~\cite{bm03} at the Thomas Jefferson
National Accelerator Laboratory in Newport News Virginia (JLab)
was divided into several running groups, each of which included
experiments that used the same target nucleus and had similar
technical needs. The G1 Running Group combined several approved
experiments that involved the interaction of real photons with the
proton. Besides the PAC approved experiments,~
\cite{wjb94,br91,rs89,jn93,gm89,rm94} there was one independent
analysis~\cite{sppc} that is
relevant to this paper. The G2 running group had many of the same
goals as the G1 running group, but used the deuteron as a neutron
target~\cite{wjb94,bm89,br94}. The G3 running group
used $^3$He and $^4$He targets to study coherent and
incoherent production of mesons and hyperon formation in
nucleus~\cite{hw91,mv93,ga93}. The G6 running group measured the
photoproduction of vector mesons at high $t$ in order to study the
region below which vector dominance is important and in which hard
processes are thought to dominate~\cite{ma93}.

\subsection{Single-Meson Photoproduction}

\begin{figure}[htbp]
\centerline{\epsfig{file=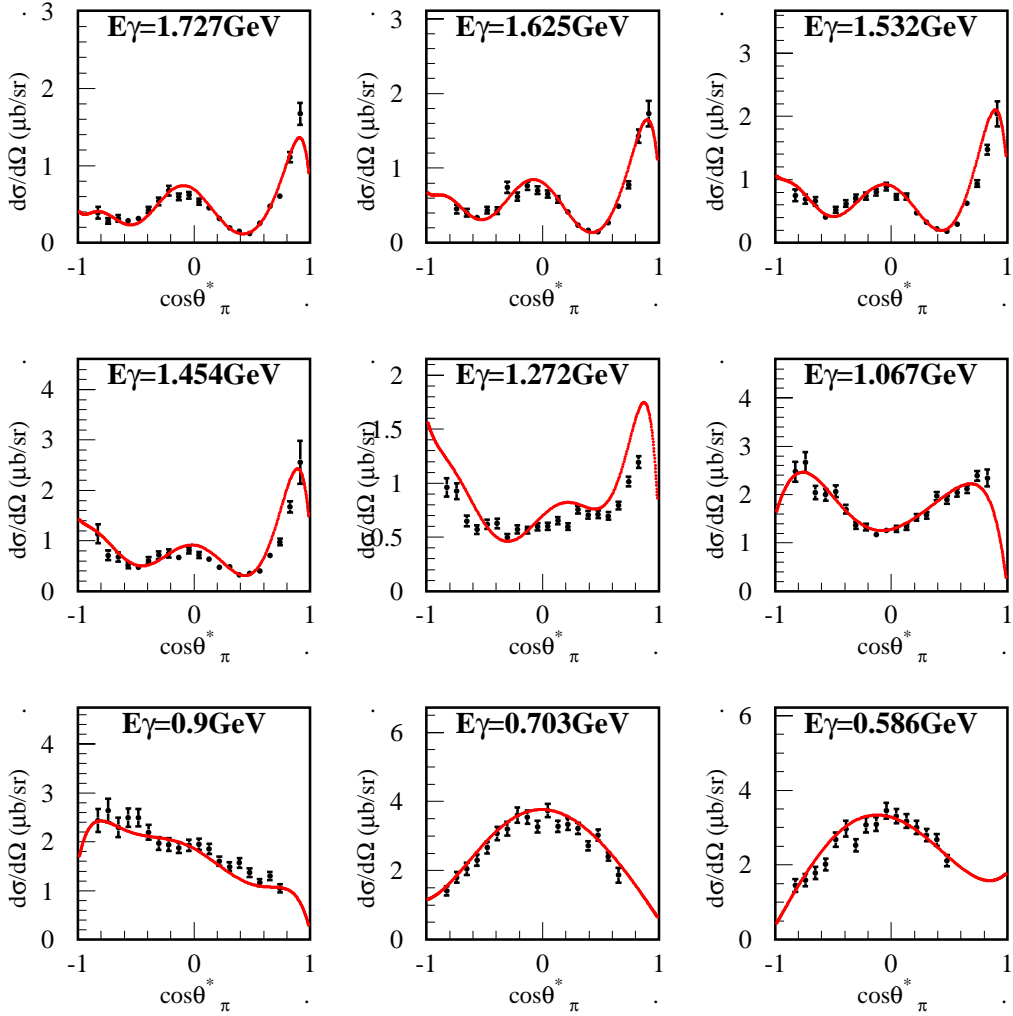,width=0.8\textwidth}}
\caption{Preliminary differential cross sections from the CLAS G1B
         run group for $\gamma p\rightarrow\pi^{\circ}p$ angular
         distributions -- arbitrary scale (solid circles)~\protect\cite{phpc}.
         These data are compared to SAID~\protect\cite{ra02}
         (solid line) predictions.} \label{fig1}
\end{figure}

Figure~\ref{fig1} shows a sample of the differential cross
sections obtained (2 $ \% $ of the G1B data). The evolution of
this set of differential cross sections is well reproduced by SAID
in the region where there is sufficient experimental data to
constrain the fit (up to 1.2~GeV).  Even up to 1.7~GeV, the
agreement is quite good.

\begin{figure}[htbp]
\centerline{\epsfig{file=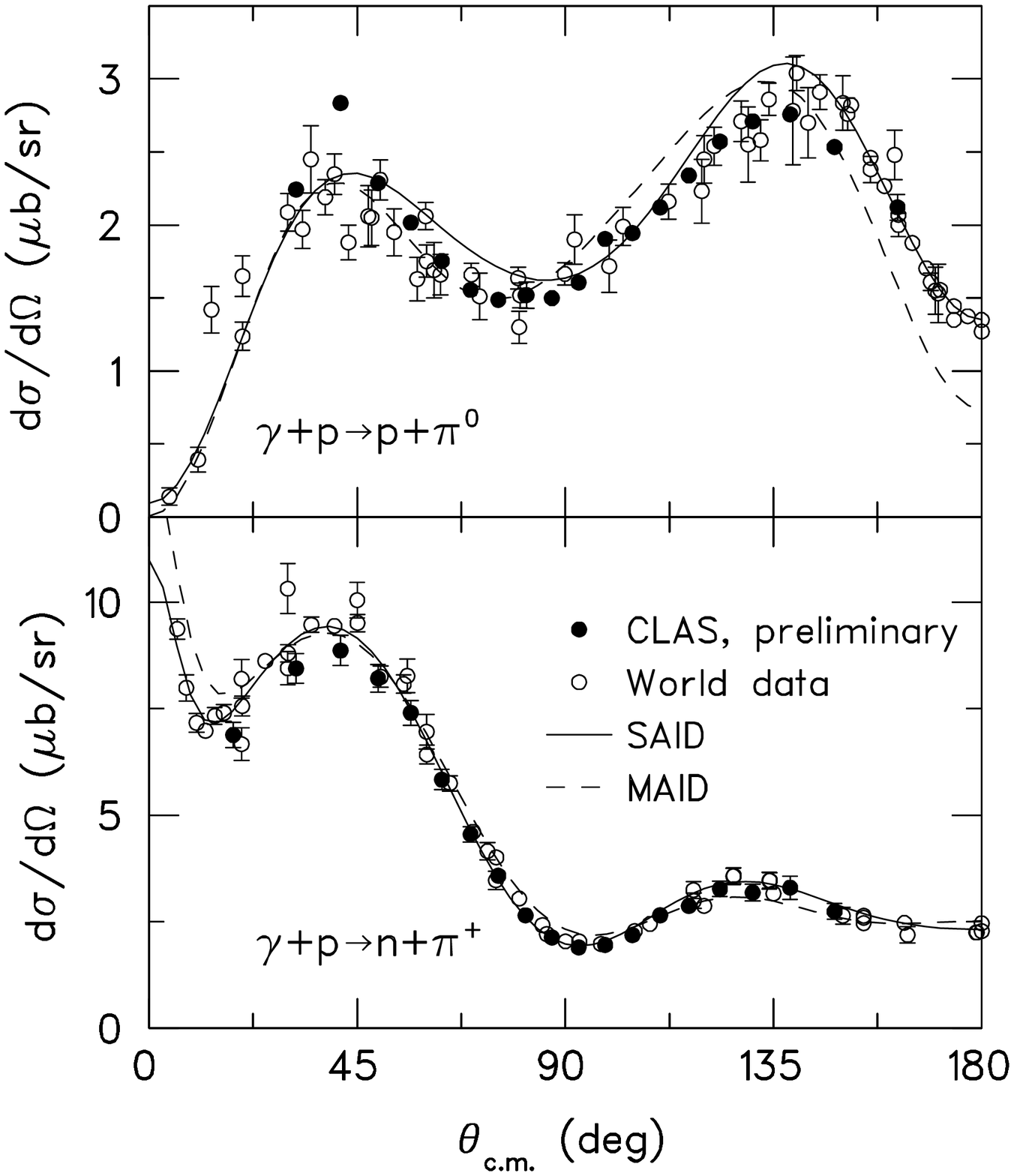,width=0.68\textwidth}}
\caption{Preliminary differential cross section data at
         $E_\gamma = 1000$~MeV from the CLAS G1C run group
         (solid circles).  These data are compared to
         other world data (open circles), along with SAID~
         \protect\cite{ra02} (solid line) and MAID~
         \protect\cite{maid} (dashed line) predictions.} \label{fig2}
\end{figure}

\begin{figure}[htbp]
\centerline{\epsfig{file=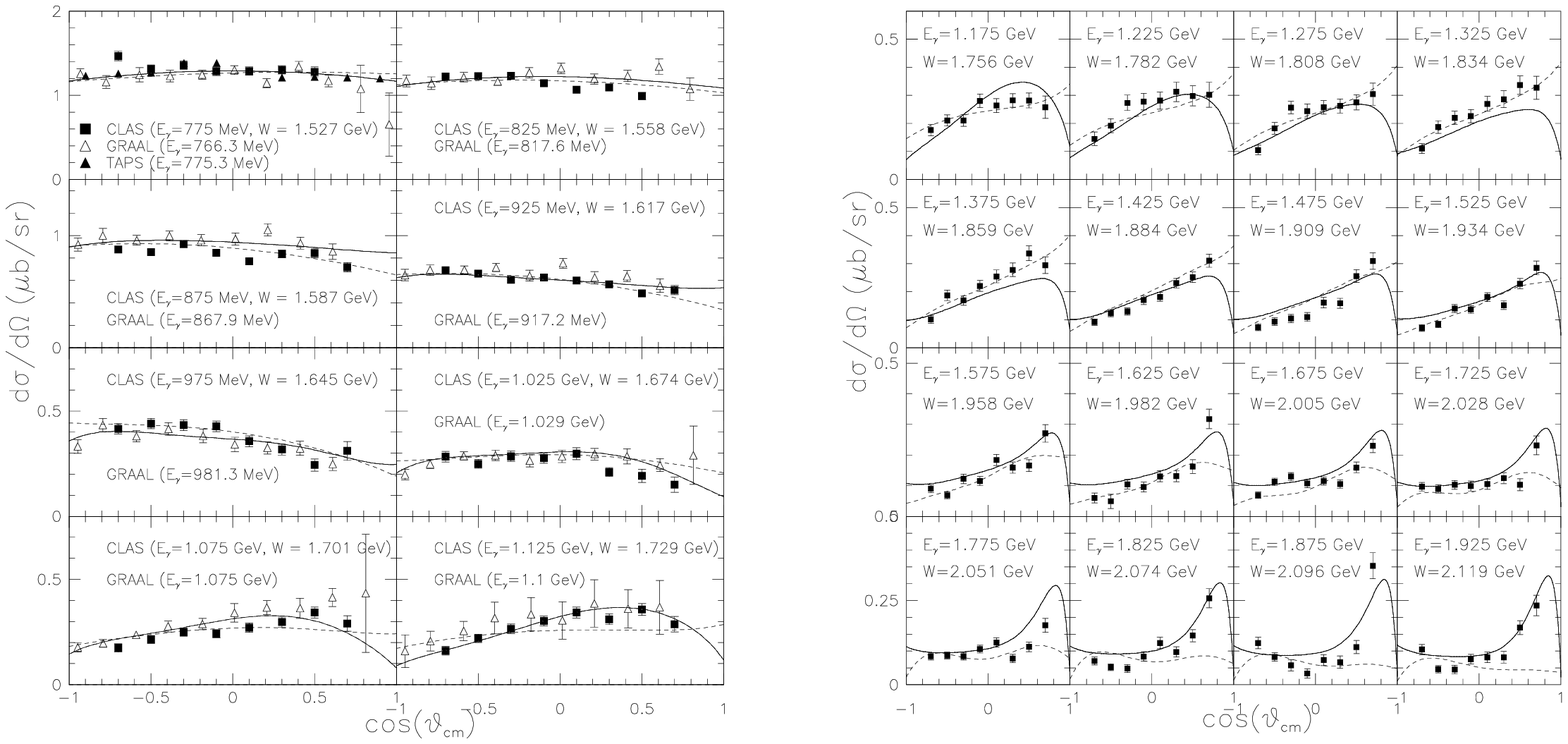,width=1.0\textwidth}}
\caption{Differential cross sections for $\gamma p\rightarrow\eta
         p$ angular distributions, 750 to 1150~MeV (left panel)
         and 1200 to 1950~MeV (right panel).  CLAS JLab data
         (solid squares)~\protect\cite{md02} are shown vs previous
         TAPS~\protect\cite{taps} and GRAAL~\protect\cite{graal}
         measurements for comparison.  Also shown are results
         from REM (solid lines)~\protect\cite{rem} and $\chi QM$
         (dashed lines)~\protect\cite{cqm} approaches.} \label{fig3}
\end{figure}

Single-pion photoproduction yields that were obtained from the G1C
data run for $\gamma p\rightarrow \pi^\circ p$ and $\gamma
p\rightarrow\pi^+n$ are shown in Fig.~\ref{fig2}.  While
normalization is still being discussed within the collaboration,
one is again comforted by the apparent good agreement between data
and SAID. Additionally, results for the reaction $\gamma
p\rightarrow\eta p$ have been obtained and are shown in
Fig.~\ref{fig3}. Here also phenomenology appears to be capable of
reproducing general features in the data.

Significant improvement in the stability of PWA solutions, over a wider
range of energies, is expected with the polarized beam and target
experiments planned for Hall B of JLab~\cite{pol_tar}. At present, double
polarization measurements are few and have little weight in a fit to
the full database.

\begin{figure}[htbp]
\hfill{}%
\includegraphics[width=0.48\textwidth,keepaspectratio]{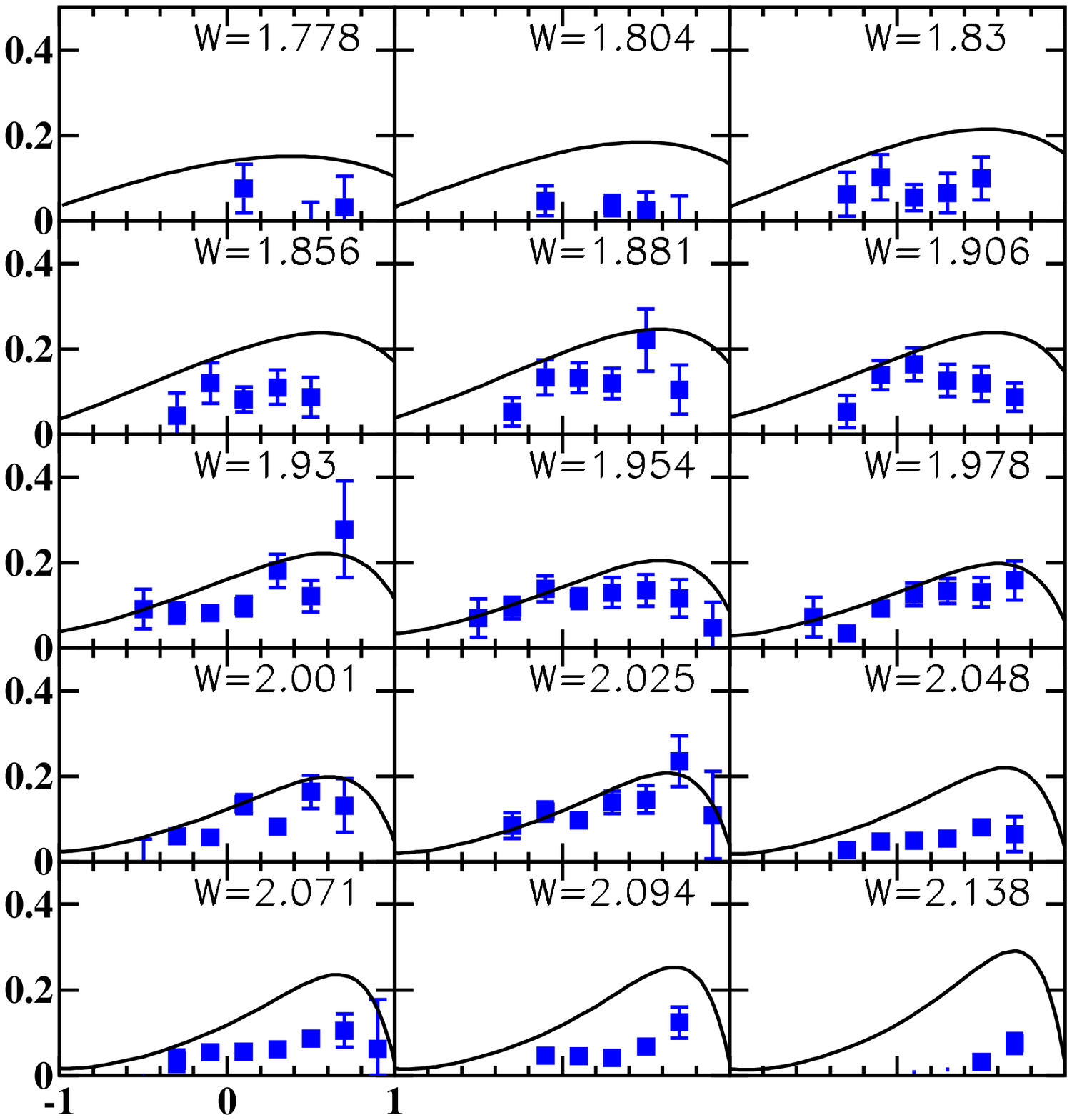}
\hfill{}%
\includegraphics[width=0.48\textwidth,keepaspectratio]{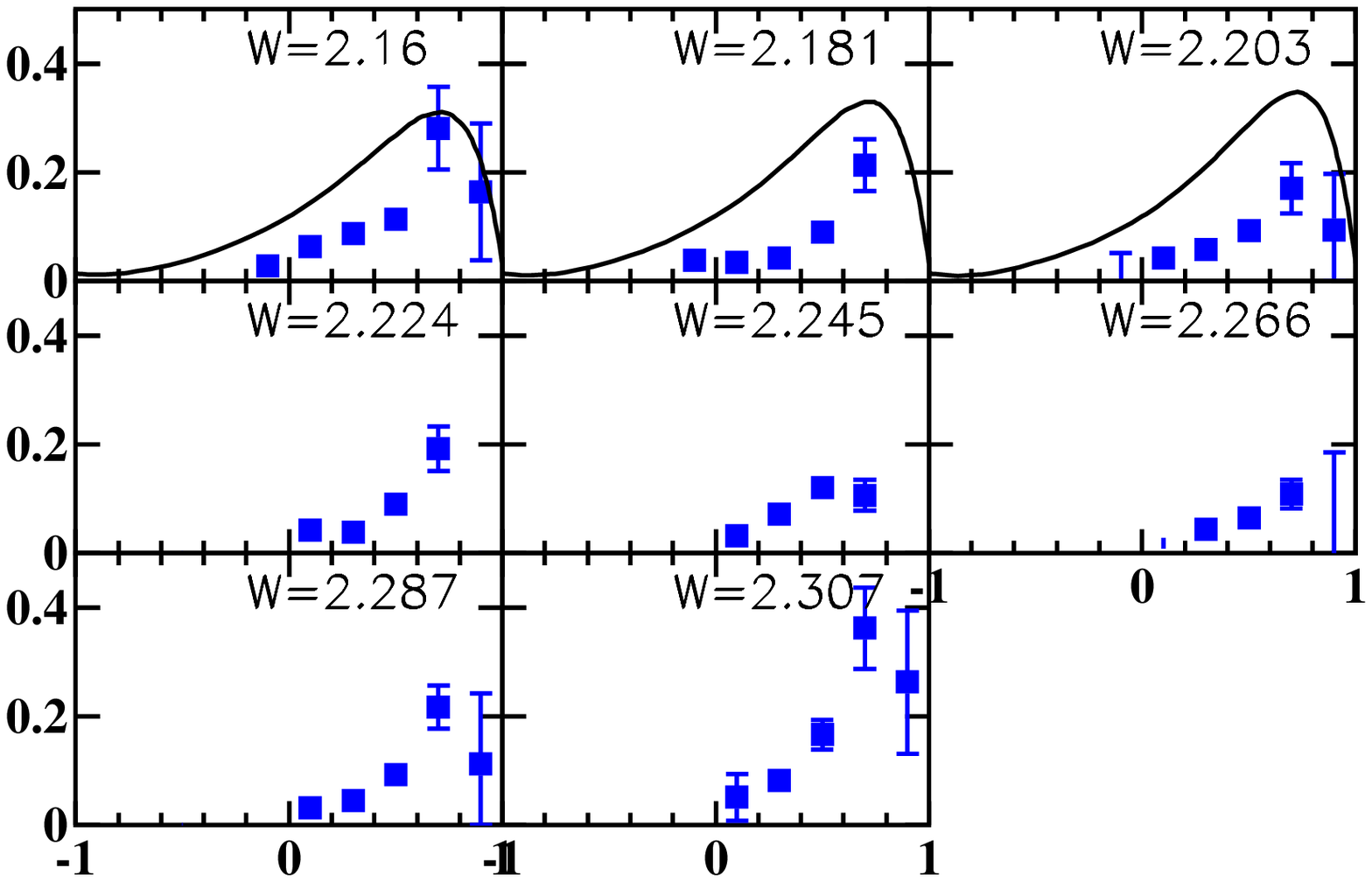}
\hfill{}%
\caption{Preliminary differential cross sections for $\gamma
         n\rightarrow K^+\Sigma^-$.  These CLAS JLab data
         ~\protect\cite{inpc} are compared to
         ~\protect\cite{Ben98} (solid line) predictions.} \label{fig4}
\end{figure}

The $\gamma n\rightarrow  K^+\Sigma^-$ channel has been
measured as part of a study of kaon photoproduction on deuterium.
The photon-energy range covered was from 0.50 to 2.95~GeV.  In the
present analysis, the reaction  $\gamma n\rightarrow  K^+\Sigma^-$
was selected by detecting the $K^+$ and the decay
products of the $\Sigma^-$. The $\pi^-$ was detected using the
time-of-flight counters and the drift chambers, while the neutron
was detected in the electromagnetic calorimeter. The neutron
momentum was determined from time-of-flight. Preliminary
differential cross sections are shown in Fig.~\ref{fig4} as a
function of the invariant energy W and the kaon polar angle in the
center-of-mass system. Here the data shows significant deviation
from a model based mainly on fits to proton-target data. These fits
are far less constrained, by both data and theory, than those associated
with the single-pion photoproduction process.

\subsection{Double-meson Photoproduction}

Double-pion photoproduction cross sections (for the $\gamma p
\rightarrow p\pi^{\circ}\pi^{\circ}$ reaction channels) extend
the study of nucleon resonances beyond previous works based upon
pion-nucleon scattering and with single-pion production.  The
$\pi\pi N$ channel will be used to extract information on
the electromagnetic excitation and decay mechanisms of resonances
at higher excitation energies, and may be essential for
disentangling the broad, overlapping resonances that do not couple
strongly to the $\pi N$  channel.  Two-$\pi^0$ production, in
particular, is not contaminated with background events from the
Born terms. Moreover, intermediate  $\rho^o(770)$ processes are
forbidden since the $\rho^o$ cannot decay into two neutral pions.

\begin{figure}[htbp]
\centerline{\epsfig{file=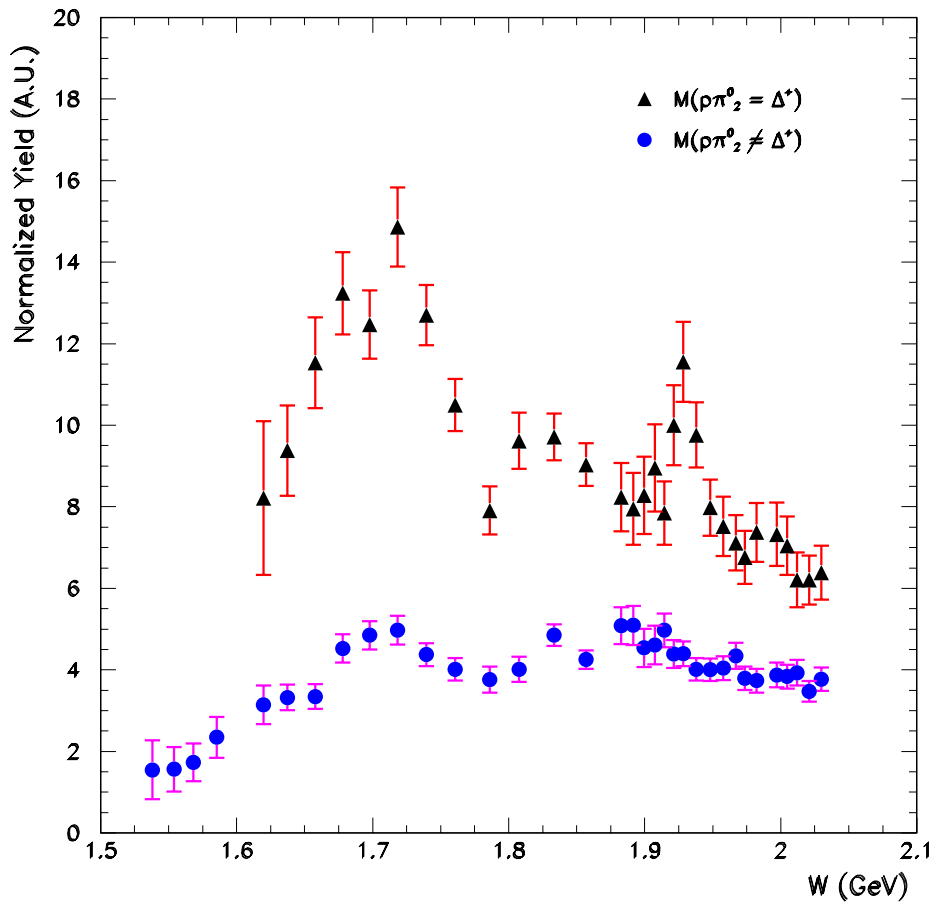,width=0.7\textwidth}}
\caption{CLAS Jlab preliminary cross sections for the reaction
         $\gamma p\rightarrow p\pi^{\circ}\pi^{\circ}$~
         \protect\cite{sppc}.} \label{fig5}
\end{figure}

Figure~\ref{fig5} shows the preliminary cross sections for the
reaction $\gamma p\rightarrow p\pi^{\circ}\pi^{\circ}$ extracted
from the two earliest G1 running periods, at photon energies
between 500 and 1700~MeV. This analysis extends the total cross
section measurements beyond $E_\gamma$ = 800~MeV. A prominent peak
at E = 1.1~GeV corresponds to a center-of-mass energy of W =
1.715~GeV; and another at E = 1.4~GeV corresponds to W = 1.9~GeV).
The cross section up to 800~MeV agrees very well with the earlier
work done at Mainz; the peak at 1.1 has been recently reported by
GRAAL but the one at 1.4~GeV has not been seen before. The general
shape of the cross section agrees with the DAPHNE measurement. It
is particularly interesting that at higher photon energies there
is evidence of some structure in the cross section, as these data
are the first of their kind in this energy region. Angular
distributions of all three final state particles will be available
once a more detailed analysis is complete. These results should
foster new theoretical work at these higher energies.

\subsection{The Crystal Ball Program at BNL/AGS}

The SLAC Crystal Ball Spectrometer was used to make measurements
at the C6 line at the Brookhaven National Laboratory, BNL,
Alternating Gradient Synchrotron, AGS, with pion momenta from 147~
MeV/c to 760~MeV/c. This series of experiments studied all neutral
final states of $\pi^-$p and $K^-$p induced reactions. New
results for the radiative capture reaction $\pi^-p\rightarrow
\gamma n$, charge exchange $\pi^-p\rightarrow \pi^{\circ}n$, two
$\pi ^{\circ}$ production $\pi^-p\rightarrow \pi^{\circ}
\pi^{\circ}n$, and $\eta$ production $\pi^-p\rightarrow \eta n$,
reactions have recently been reported. Data were taken
simultaneously on all reactions, which ensures that background
events were accurately subtracted. Data taking using the Crystal
Ball began in July 1998 and continued until late November 1998. An
additional two-week run was completed in May 2002, just before
disassembling the Crystal Ball for shipment to Mainz.

The Crystal Ball is a segmented, electromagnetic calorimetric
spectrometer, covering 94\% of $4\pi $ steradians. It was built at
SLAC and used for meson spectroscopy measurements there for three
years. It was then used at DESY for five years of experiments and
put in storage at SLAC from 1987 until 1996 when it was moved to
BNL by our collaboration. The Crystal Ball is constructed of 672
hygroscopic NaI crystals, hermetically sealed inside two
mechanically separate stainless steel hemispheres. Each crystal is
viewed by a photomultiplier tube, PMT. There is an entrance and
exit tunnel for the beam, LH$_{2}$ target plumbing, and veto
counters. Each crystal is shaped like a truncated triangular
pyramid, points toward the interaction point, is optically
isolated, and is viewed by a PMT which is separated from the
crystal by a glass window. The beam pipe is surrounded by 4
scintillators covering 98\% of the target tunnel and form the
veto-barrel. Electromagnetic showers are measured with an energy
resolution of $\Delta$E/E = 0.02/[E(GeV)]$^{0.36}$, and an angular
resolution for $\theta$ of $2^\circ$--$3^\circ$, for energies in the
range 50--500~MeV, and a resolution in $\phi$ of $2^\circ/{ \sin
\theta}$. The energy calibration is done \textit{in situ} using
the reactions: i) $\pi^-p\rightarrow\gamma n$\ at rest,
yielding an isotropic, monochromatic $\gamma$ flux of 129.4~MeV;
ii) $\pi^-p\rightarrow\pi^{\circ}n$\ at rest, yielding a
pair of photons in the energy range 54.3--80~MeV, almost back to
back; and iii) $\pi^-p\rightarrow\eta n$\ {\ }at threshold,
yielding two photons, about 300~MeV each, in coincidence almost
back to back. Scintillators surround the LH2 target to provide a
charged particle veto and a beam veto scintillator is located
downstream of the target as is a Cerenkov counter to monitor
electron contamination in the beam. The trigger consists of: a
beam coincidence trigger, no downstream beam veto, and a total
energy-over-threshold signal from the Crystal Ball. A trigger
based on the distribution of the energy in different regions of
the Crystal Ball was also used to provide a more restrictive
trigger in certain cases.

\subsubsection{The Radiative Decay of Nucleon Resonances}

The radiative decay of a resonance provides an excellent
laboratory for testing theories of the strong interaction,
allowing us to probe the structure of the nucleon itself. In
particular, radiative-capture data are important in the study of
the poorly understood neutral Roper resonance. They can be
combined with recent JLab Hall B data for the reactions $\gamma
p\rightarrow \pi^+n$ and $\gamma p\rightarrow\pi^{\circ}$p, which
have contributions from mesonic decays of the charged Roper,
excited by incident photons with energies from 400 to 700~MeV. In
addition, comparison of radiative capture data to the new JLab
data taken in the inverse reaction $\gamma n \rightarrow \pi^-p$,
using a deuteron target, tests extrapolation techniques for the
deuteron correction and allows one to study medium effects within
the deuteron. Figure 6 shows the results of our radiative capture
measurements at the equivalent $E_\gamma=285~MeV$ compared the
latest SAID\cite{ra02} and MAID\cite{maid} differential cross
section predictions. On average, over all 18 energies measured at
BNL, the new data favor the SAID predictions. One see very little
change in cross section when our data are included in the SAID
fit. However, there are some hints of significant changes in the
electric and especially the magnetic multipoles.

\begin{figure}[htbp]
\centerline{\epsfig{file=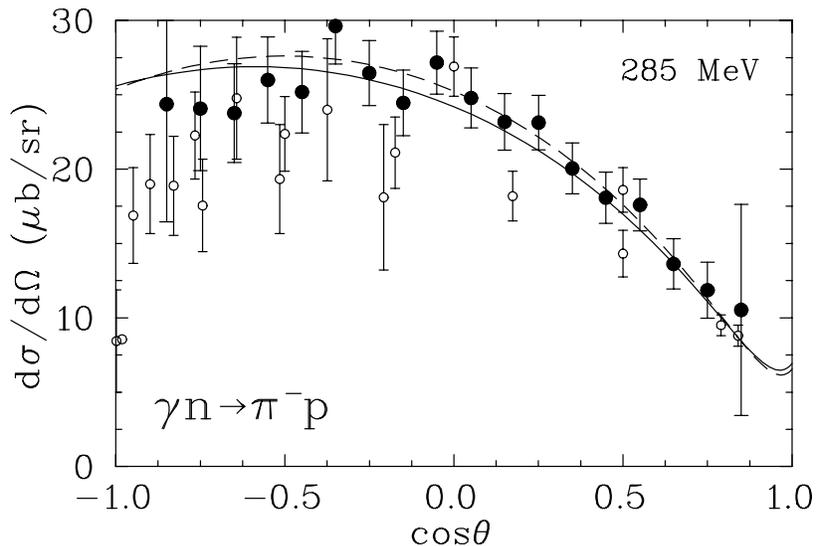,width=0.4\textwidth,angle=90.0}}
\caption{Preliminary differential cross section data
         for $\gamma n\rightarrow\pi^-p$ at
         $E_\gamma = 285$~MeV from the Crystal Ball
         (solid circles)~\protect\cite{aziz}.  These data are compared to
         other world data (open circles), along with SAID~
         \protect\cite{ra02} (solid line) and MAID~
         \protect\cite{maid} (dashed line) predictions.} \label{fig6}
\end{figure}

The primary reason that so few data are available for the
radiative capture reaction is the difficulty in separating its
contribution from other reactions. This is mainly due to the
significant background from $\pi^-p\rightarrow\pi^{\circ}n$\ whose
cross-section is about 50 - 100 times larger. The geometry of the
Crystal Ball provides the capability of discriminating against
multiple $\gamma $-rays that arise from the decay of $\pi
^{\circ}$s and $\eta$s. However, because of the large entrance and
exit tunnels, there is a 20\% chance that one of the two $\gamma
$'s from say $\pi^{\circ}$ decay is missed, resulting in a false
one-photon event. The separation of signal for
$\pi^-p\rightarrow\gamma n$\ from ``background" was investigated
using GEANT and EHS. Our Monte-Carlo includes such subtleties as
secondaries from photon and pion breakup of a nucleus in the NaI,
photon split-offs (a single photon cluster split into two) and
backward Compton scattering. It reproduces and improves upon the
photon energy and angular resolution that were measured in the
course of the Crystal Ball's eight-year tenure at SLAC and DESY.
The large solid angle acceptance of the Crystal Ball  lead to a
rejection factor of about 40--150 for the background events from
$\pi^-p\rightarrow\pi^{\circ}n$\ .

\subsubsection{Charge Exchange}

The charge-exchange process has been the weakest link in $\pi N$
partial-wave and coupled-channel studies. The accurate data that
we have obtained in this momentum region will help in improving
the determinations of the isospin-odd s-wave scattering length,
the $\pi$NN coupling constant, and the $\pi$-N $\sigma$ term. In
addition, better charge exchange data helps in evaluating the mass
splitting of the $\Delta$ resonance and may result in new values
for the P$_{11}$(1440) mass and width. Figure 7 shows the charge
exchange data used in the evaluation of background in our crystal
ball measurements. A full reanalysis using these data is in
progress. The solid line representing the current SAID fit does
not include these data.

\begin{figure}[htbp]
\centerline{\epsfig{file=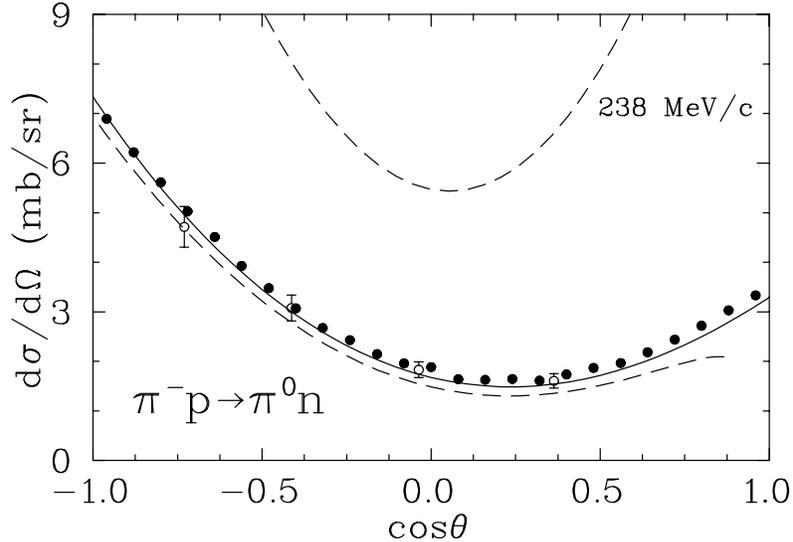,width=0.4\textwidth,angle=90.0}}
\caption{Preliminary differential cross section data
         for $\pi^{-}p\rightarrow\pi^{\circ}n$ at
         $E_\gamma = 238$~MeV/c from the Crystal Ball
         (solid circles).  These data are compared to
         other world data (open circles), along with SAID~
         \protect\cite{fa02} (solid line) predictions.
         Dashed lines associated with the triangle inequality.}
         \label{fig7}
\end{figure}

\subsubsection{Near-Threshold Eta Production}

\begin{figure}[htbp]
\centerline{\epsfig{file=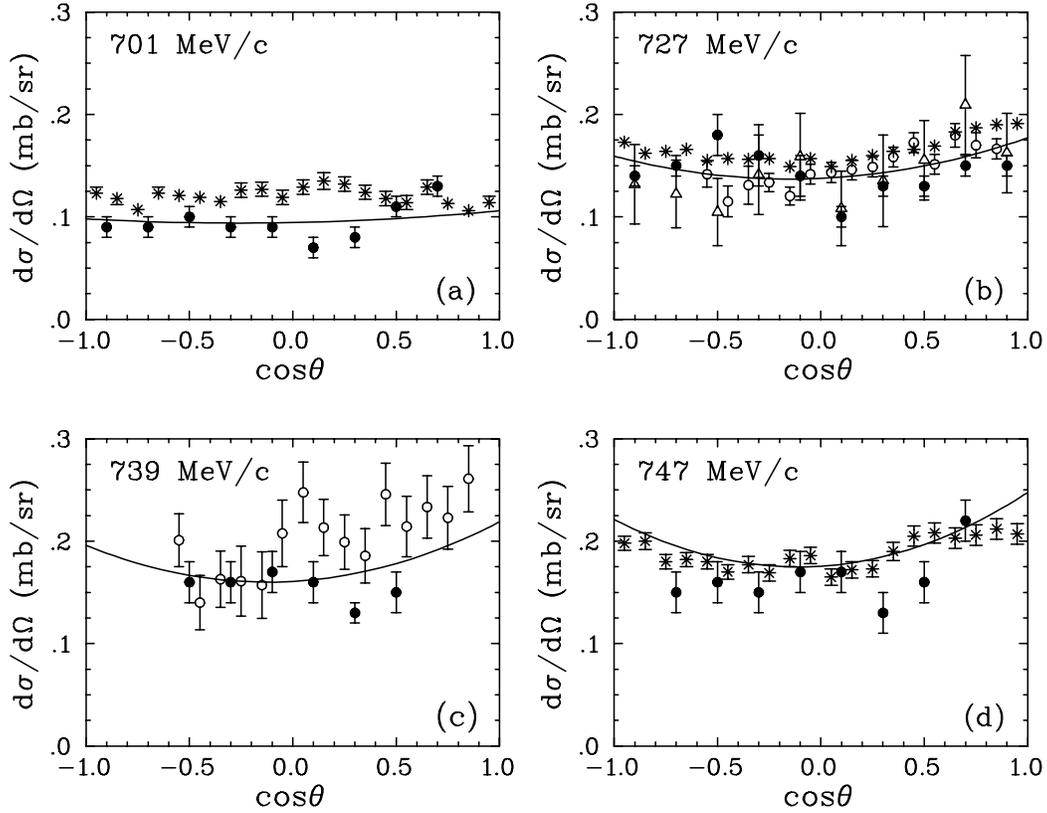,width=0.6\textwidth,angle=90.0}}
\caption{Preliminary differential cross section data
         for $\pi^{-}p\rightarrow\eta n$ at from E909 of
         BNL/AGS~\protect\cite{tom} (solid circles) and the Crystal
         Ball E913/914 BNL/AGS~\protect\cite{ko03} (stars).  These
         data are compared to other world data (open circles),
         along with SAID~\protect\cite{fa02} (solid line) predictions.} \label{fig8}
\end{figure}

Near-threshold $\eta$-production measurements provide data useful
in verifying models of $\eta$-meson production and are necessary
for an extraction of the $\eta$-N scattering length. They will
also be necessary to resolve ambiguities in the resonance
properties of the S$_{11}$(1535) and in the $\eta$ photoproduction
helicity amplitudes. In Fig.~\ref{fig8}, low-energy data are
compared to a recent GW coupled-channel fit~\cite{fa02} to $\pi N$
elastic and $\pi^- p \to \eta p$ data. Here one can see the level
of conflict between datasets, and the much improved statistical
uncertainties associated with the Crystal Ball data.\cite{ko03} It
should be pointed out that the lowest energy points for the
crystal ball data set is a few MeV higher than those of Morrison
\textit{et al.}\cite{tom} which were take with a small acceptance
detector previous to the arrival of the crystal ball at BNL.
Accounting for the higher energy and the steep rise in cross
section at threshold, one would expect a somewhat higher value
that is still fairly flat indicating a mostly S-wave behavior very
near threshold.

\subsection{Summary and Conclusions}

The baryon spectroscopy programs at JLab and BNL have produced a
plethora of data points that enhance our ability to do new
partial-wave analyses. Besides the single-channel analyses of the
past, several coupled channel analyses are being accomplished
owing to the expansion of the data base. It is anticipated that
final results of the experiments are eminent and that our new fits
can be firmed up. With the apparent "discovery" of the pentaquark,
a revival of "N-Star" physics seems very likely with even the most
unlikely candidates jumping on the bandwagon. It is hoped that
this resurgence of interest will allow those who have been long
standing workers in the field to continue their efforts and reap
the fruits of their labors. One word on caution - the analyses in
progress must be fully and carefully completed as expeditiously as
possible. Without these data the data base remains incomplete.

\section{Acknowledgments}

The author acknowledges the support of the United States
Department of Energy (DOE), The Southeastern Universities Research
Association (SURA), The Thomas Jefferson National Accelerator
Facility (JLab), and the George Washington University Research
Enhancement Fund.




\end{document}